\PassOptionsToPackage{pdfpagelabels=false}{hyperref}
\documentclass[fleqn,usenatbib,usedcolumn]{mnras}
\usepackage[british]{babel}             

\usepackage{newtxmath,newtxtext}
\DeclareSymbolFont{operators}{OT1}{ntxtlf}{m}{n}
\SetSymbolFont{operators}{bold}{OT1}{ntxtlf}{b}{n}

\usepackage[T1]{fontenc}

\usepackage{graphicx}    
\usepackage{amsmath}    
\usepackage{amssymb}    
\usepackage{xspace}
\usepackage{xstring}
\usepackage{pifont}
\usepackage{pgfkeys}

\newcommand{\rv}{\ensuremath{v_{\textrm{rad}}}}
\newcommand{\ofe}{\ensuremath{[\textrm{O}/\textrm{Fe}]}}
\newcommand{\nafe}{\ensuremath{[\textrm{Na}/\textrm{Fe}]}}

\newcommand{\feh}{\ensuremath{[\textrm{Fe}/\textrm{H}]}}
\newcommand{\bafe}{\ensuremath{[\textrm{Ba}/\textrm{Fe}]}}
\newcommand{\alfe}{\ensuremath{[\textrm{Al}/\textrm{Fe}]}}
\newcommand{\mgfe}{\ensuremath{[\textrm{Mg}/\textrm{Fe}]}}
\newcommand{\yfe}{\ensuremath{[\textrm{Y}/\textrm{Fe}]}}
\newcommand{\alphafe}{\ensuremath{[\upalpha/\textrm{Fe}]}}

\newcommand{\ebv}{\ensuremath{\textrm{E}(\textrm{B}-\textrm{V})}}

\newcommand{\teff}{\ensuremath{\textrm{T}_\textrm{eff}}\xspace}
\newcommand{\logg}{\ensuremath{\log \textrm{g}}\xspace}
\newcommand{\kms}{\ensuremath{\textrm{km}\,\textrm{s}^{-1}}}

\newcommand{\masyr}{\ensuremath{\textrm{mas}\,\textrm{yr}^{-1}}}
\newcommand{\pmra}{\ensuremath{\mu_\textrm{RA}}}
\newcommand{\pmdec}{\ensuremath{\mu_\textrm{Dec}}}
\newcommand{\ocen}{$\upomega$~Cen\xspace}
\newcommand{\bprp}{\ensuremath{\textrm{G}_\textrm{\textsc{bp}}-\textrm{G}_\textrm{\textsc{rp}}}}
\defcitealias{Ibata2019a}{I+19}
\newcommand{\gaia}{{\it Gaia}\xspace}
\newcommand{\sourceid}[1]{%
    \StrLeft{#1}{1}[\firstletter]%
    \StrRight{#1}{3}[\lastletter]%
    \firstletter...\lastletter\xspace
}
\newcommand{\response}[1]{{#1}}

\title[GALAH Survey: Chemically tagging streams to \ocen]{The GALAH Survey: Chemically tagging the Fimbulthul stream to the globular cluster $\upomega$ Centauri}

\author[J. D. Simpson et al.]{
\parbox{\textwidth}{\raggedright
Jeffrey~D.~Simpson,$^{1}$\thanks{Email: \texttt{jeffrey.simpson@unsw.edu.au}}
Sarah~L.~Martell,$^{1,2}$
Gary~Da~Costa,$^{3}$
Jonathan~Horner,$^{4}$
Rosemary~F.~G.~Wyse,$^{5}$
Yuan\nobreakdash-Sen~Ting,$^{6,7,8}$
Martin~Asplund,$^{2,3}$
Joss~Bland\nobreakdash-Hawthorn,$^{2,9}$
Sven~Buder,$^{10}$
Gayandhi~M.~De~Silva,$^{2,11}$
Ken~C.~Freeman,$^{3}$
Janez~Kos,$^{9,12}$
Geraint~F.~Lewis,$^{9}$
Karin~Lind,$^{10,13}$
Sanjib~Sharma,$^{2,9}$
Daniel~B.~Zucker,$^{2,14}$
Toma\v{z}~Zwitter,$^{12}$
Klemen~\v{C}otar,$^{12}$
Peter~L.~Cottrell$^{15,16}$
and Thomas~Nordlander$^{2,3}$
\\
{\small{\it(Affiliations are listed after the references)}}}}

\date{Accepted 2019 November 4. Received 2019 October 20; in original form 2019 August 6}

\pubyear{2019}

\begin{document}
\label{firstpage}
\pagerange{\pageref{firstpage}--\pageref{lastpage}}
\maketitle

\begin{abstract}
Using kinematics from \gaia and the large elemental abundance space of the second data release of the GALAH survey, we identify two new members of the Fimbulthul stellar stream, and chemically tag them to massive, multi-metallic globular cluster $\upomega$ Centauri. Recent analysis of the second data release of \gaia\ had revealed the Fimbulthul stellar stream in the halo of the Milky Way. It had been proposed that the stream is associated with the \ocen, but this proposition relied exclusively upon the kinematics and metallicities of the stars to make the association. In this work, we find our two new members of the stream to be metal-poor stars that are enhanced in sodium and aluminium, typical of second population globular cluster stars, but not otherwise seen in field stars. Furthermore, the stars share the s-process abundance pattern seen in \ocen, which is rare in field stars. Apart from one star within 1.5~deg of \ocen, we find no other stars observed by GALAH spatially near \ocen\ or the Fimbulthul stream that could be kinematically and chemically linked to the cluster. Chemically tagging stars in the Fimbulthul stream to \ocen\ confirms the earlier work, and further links this tidal feature in the Milky Way halo to \ocen.
\end{abstract}



\section{Introduction}\label{sec:intro}
The modern paradigm for the formation and growth of galaxies, and of larger scale structures within the Universe (known as the $\Lambda$CDM paradigm) features the growth of large galaxies, like the Milky Way, through mergers with smaller bodies \citep[e.g.,][]{Searle1978,White1991,Cote1998,Cote2000,Venn2004,Cole2002,Oser2010}. When a dwarf galaxy merges with Milky Way, it is tidally stripped of some, or all, of its stars. Similarly, objects already orbiting the Milky Way can be tidally stripped of stars during events like Galactic plane passages. If the dynamical evolution timescale is larger than the orbital period, these stripped stars remain for some time on approximately the mean orbit of the progenitor \citep[][]{Lynden-Bell1995}. This leads to streams and tidal debris throughout the Milky Way Galaxy, acting as the fossils records of these events.

Large-scale photometric surveys have, to date, uncovered some 50 streams in our Galaxy \citep[][and references therein]{Mateu:2017}. The second data release of the ESA's \gaia mission \citep{GaiaCollaboration:2016cu,GaiaCollaboration:2018io} has caused an explosion in the discovery rate of streams through the combination of precise photometry and proper motion measurements, and has been used to identify stellar streams in the local region of the Galaxy \citep{Meingast2019}, the bulge \citep[e.g.,][]{Ibata:2018wn}, the halo \citep[e.g.,][]{Malhan2018,Koppelman2018,Ibata2019}, and even potentially outside of our Galaxy in the Magellanic Stream \citep[e.g., Price-Whelan 1;][]{Price-Whelan2018,Nidever2019a,Bellazzini2019}. There is much interest in kinematically mapping these streams \citep[e.g.,][]{Li2019a} as they are a useful tool for studying our Galaxy's history and large-scale structure \citep{Eyre2009,Law2010,Bonaca2018,Malhan2018a}.

A number of extant star clusters have been found to have extra-tidal features or streams associated with them \citep[e.g.,][]{Odenkirchen2001, Belokurov2006a, Olszewski2009, Kuzma2018, Palau2019, Carballo-Bello2019}. Such links are particularly valuable, as they can provide a strong test for models of the gravitational field of the Milky Way. Simply put, any successful model must correctly predict the location and motion of the stream relative to the progenitor cluster.

\citet{Ibata2019} identified a new cluster-stream pairing using \gaia DR2, finding that the massive globular cluster \ocen could be linked to one of their newly discovered streams, which they named Fimbulthul. They found that Fimbulthul and \ocen had similar orbital energies and angular momenta. The relationship was explored further in \citet[][hereafter \citetalias{Ibata2019a}]{Ibata2019a}, where the authors estimated the metallicities of four stream members, finding them to be metal poor. Since \citetalias{Ibata2019a} had only metallicities for their stars to compare to \ocen, and \ocen\ has a large metallicity range \citep[$-2.0<\feh<-0.5$; e.g.,][]{Johnson:2010fs}, it is prudent to explore a larger set of elemental abundance distributions of the stream stars, and to determine if they can be chemically tagged to \ocen. The chemical tagging of stars in streams to progenitors is a burgeoning field of research. \citet{Hasselquist2019} found $\sim30$ stars in the APOGEE survey that had kinematics consistent with those predicted by N-body simulations of the Sagittarius dwarf spheroidal galaxy stream and chemical abundances like those found in it.

In general, chemically tagging stars to globular clusters takes advantage of the fact that almost\footnote{The notable exception is Ruprecht 106 \citep{Villanova:2013fk,Dotter2018}.} every Milky Way globular cluster studied in detail exhibits star-to-star variations in light-element abundances\footnote{The formation mechanisms of globular clusters are hotly debated; see the review of \citet{Bastian:2018kj} and references therein.}, typically with anti-correlations between \ofe\ \& \nafe, and \mgfe\ \& \alfe. In globular cluster research, the stars with abundance patterns like the Galactic halo (e.g., low sodium and high oxygen) are commonly referred to as `primordial' or `first population' stars. The stars with enhanced abundances of sodium and depleted oxygen are `enriched' or `second population' stars. It is these second population stars that can be most readily identified in the field \citep[as they are only about $\sim2$ per cent of the halo, e.g.,][]{Martell:2011ej} and chemically tagged as having come from globular clusters \citep[e.g.,][]{Martell:2010is,FernandezTrincado:2016db,Fernandez-Trincado2017,Schiavon:2017dg,Tang2018}.

In addition to the light element abundance dispersions, \ocen\ exhibits large metallicity and s-process element abundance ranges \citep[e.g.,][]{Norris1995,Marino:2012eg,Johnson:2010fs,Simpson:2012kc,Simpson:2013ee}, neither of which are typically observed in globular clusters \citep[there are some notable exceptions, e.g., M22;][]{Marino2011a}. Along with its retrograde orbit about the Milky Way, these abundance patterns have led to the hypothesis that \ocen\ is the core of an accreted dwarf galaxy \citep{Bekki2003}. If so, there should be a large amount of tidal debris in the Milky Way halo that was stripped from \ocen's parent galaxy.

There have been many searches for such stars. The obvious place to search for tails is within 10~deg of the cluster, where the kinematic and distance properties will still be similar to those of the cluster. \citet{Marconi2014} and \citetalias{Ibata2019a} both find evidence for tails in star counts. While those photometric surveys have provided some evidence for extra-tidal features, spectroscopic searches within a few degrees have not identified significant quantities of \ocen\ stars beyond the tidal radius. \citet{DaCosta2008} searched up to 2~deg from the cluster and found only six stars outside the tidal radius with radial velocities expected for \ocen\ --- and we note that with \gaia proper motions only two of these are consistent with membership.

A few field stars have been associated with \ocen. \citet{Lind:2015gz} found a star with low \mgfe\ and high \alfe\ that they conjectured could have been ejected at high speed from \ocen. \citet{Fernandez-Trincado2015} found eight stars in RAVE that they calculated to have been ejected from \ocen. Both of these works warrant follow-up in light of the new precision proper motions from \gaia. \citet{Myeong2018} identified retrograde substructures in the halo based upon clustering in action space, and some of these structures they associated with \ocen.

In this work we aim to chemically tag stars of the Fimbulthul stream to \ocen\ using the second data release of the GALactic Archaeology with HERMES (GALAH) survey \citep{DeSilva:2015gr,Buder2018}. The GALAH survey is a large and ambitious spectroscopic investigation of the local stellar environment. One of its principal aims is to determine precise abundances of up to 30 elements from one million stars, and to use chemical tagging to identify dispersed stellar clusters in the field of the disk and halo \citep[for the initial motivating chemical tagging papers, see][]{Freeman:2002kz,BlandHawthorn:2010kt}. The results from the GALAH survey have already been used to explore stellar streams \citep{Quillen:2018fi}, have shown that chemical tagging can work on co-moving pairs of stars \citep{Simpson:2018wu}, and have found additional members of the Pleiades \citep{Kos:2018fy}. It has also had the interesting null result of showing that several purported clusters do not exist \citep{Kos:2018we}.

This paper is structured as follows: we describe the datasets used in Section~\ref{sec:data}, the spatial and kinematic identification of stream stars in Section~\ref{sec:stream_ident}, the chemical tagging of the stars to \ocen\ in Section~\ref{sec:chemical_tagging}, and we present our conclusions in Section~\ref{sec:discussion}.

\section{Data}\label{sec:data}
In this work, we use the second public data release of the GALAH survey \citep[GALAH DR2; see the release paper:][]{Buder2018}, as well as \ocen\ stars observed by GALAH that were not part of the public release. Overall, this gives 375,688 stars with up to 23 elemental abundances per star. The GALAH survey's observing procedures can be found in \citet{Martell:2017jd}. In-depth descriptions of the spectral reduction, stellar parameter and abundance inference pipelines can be found in \citet{Kos:2017es} and \citet{Buder2018}. Below we will briefly describe the details from those papers pertinent to this work.

These data are based upon spectra obtained between 2013 November and 2017 September using the 3.9-metre Anglo-Australian Telescope with the HERMES spectrograph \citep{Sheinis:2015gk} and the Two-Degree Field fibre positioner top-end \citep{Lewis:2002eg}. HERMES simultaneously acquires spectra for $\sim360$ science targets using four independent cameras with non-contiguous wavelength coverage totalling $\sim1000$~\AA\ at a spectral resolving power of $R\approx28,000$. The spectra were reduced using an \textsc{iraf}-based pipeline that was developed specifically for GALAH and optimized for speed, accuracy, and consistency \citep{Kos:2017es}.

The GALAH DR2 stellar parameter and abundance pipeline used a two-step process \citep{Buder2018}. In the first step, spectra with high signal-to-noise were identified and analyzed with the spectrum synthesis code Spectroscopy Made Easy \citep[\textsc{sme};][]{Valenti:1996hf, Piskunov:2016ej} to determine the stellar labels (e.g., \teff, \logg, \feh, $v_\textrm{mic}$, $v\sin i$, $v_\textrm{rad}$, and [X/Fe]). In the second step, \textsc{the cannon} \citep{Ness:2015kv} learned these training set labels from \textsc{sme} and built a quadratic model at each pixel of the normalized spectrum as a function of the labels. Abundance estimates were then generated from \textsc{the cannon} model for all of the spectra. It is these \textsc{the cannon} values that we use in this work.

For some stars, the label results from \textsc{the cannon} are not necessarily reliable. For stellar parameters, these are flagged in the \texttt{flag\_cannon} value, and for individual elements in the \texttt{flag\_x\_fe} value (where \texttt{x} is the element symbol). \citet{Buder2018} recommends to only use stellar parameters from stars where $\texttt{flag\_cannon}=0$, and for a given elemental abundance where $\texttt{flag\_x\_fe}=0$. However, this would restrict the number of possible \ocen-like stars we could investigate, as the training set was not highly populated at the metal-poor end of the iron abundance distribution function. Of the 9090 stars in the dataset with $\feh<-1.0$, only 66 per cent (5987/9090) have $\texttt{flag\_cannon}=0$, and 1929 stars have $\texttt{flag\_cannon}=1$ (this indicates that \textsc{the cannon} started to extrapolate outside of the training set bounds to estimate the abundance). For the elements of interest for chemical tagging stars to \ocen, many of the metal-poor stars would be excluded with a strict flagging criterion: for \nafe\ and \bafe\ only $\sim2700$ stars of these metal-poor stars have $\texttt{flag\_x\_fe}=0$, and for \alfe\ it is only 628 stars. In this work, we decided to loosen the flagging criteria and use iron abundance values for stars where $\texttt{flag\_cannon}\in\{0,1\}$ (i.e., no flags; or \textsc{the cannon} started to extrapolate) and individual elemental abundances where $\texttt{flag\_x\_fe}\in\{0,1,2\}$ (i.e., no flags; \textsc{the cannon} started to extrapolate; or line strength was below the 2-$\sigma$ upper limit). This increases the number of useful stars to 7916, 4326, 5225, and 1886 for \feh, \nafe, \bafe, and \alfe\ respectively.

The GALAH survey data was combined with the astrometric and proper motion data from the second data release (DR2) from the \gaia mission. GALAH's input catalogue is based upon 2MASS \citep{Skrutskie:2006hl}, so we used the \texttt{tmass\_best\_neighbour} table produced by the \gaia collaboration for the cross-match. To exclude stars with likely poor astrometry, we required that the Renormalised Unit Weight Error \citep[defined in][]{Lindegren2018} be $<1.4$. As \gaia photometry is less reliable in crowded regions (e.g., globular clusters) due to source confusion in the blue and red photometers \citep{Evans:2018cj}, we excluded stars which did not fall in the range $1.0 + 0.015(\bprp)^2<\texttt{CE}<1.3 + 0.06(\bprp)^2$, where \texttt{CE} is the \texttt{phot\_bp\_rp\_excess\_factor} \citep{Babusiaux:2018di}.

From both the GALAH dataset and the full \gaia dataset we identified likely \ocen members. For both datasets we selected stars within 1~deg of $(\textrm{RA},\textrm{Dec})=(201.697\degr,-47.480\degr)$, proper motions within 2~\masyr\ of $(\pmra,\pmdec)=(-3.1925,-6.7445)$~\masyr\ \citep{Helmi:2018dl}, and parallaxes of $|\varpi|<1$~mas. The proper motion distance of $<2$~\masyr\ was selected as it encompassed an obvious over-density in the proper motion distribution of the GALAH-observed stars. The systemic radial velocity of \ocen has a large offset from the bulk of stars along the line-of-sight: 232~\kms\ \citep[see, e.g.,][]{DaCosta2008}. For the GALAH dataset we required that every star be within 40~\kms\ of this value. For the \gaia dataset, the bulk of stars do not have measured radial velocities, so we excluded only the few stars that met the other criteria but were outside the $232\pm40$~\kms\ range. In total, these selection criteria identified 352 members of \ocen\ observed by the GALAH survey (with a radial velocity mean and standard deviation of $232.9\pm8.6$~\kms), and 66,597 from \gaia DR2.

\section{Identification of new candidate stream members}\label{sec:stream_ident}

\begin{table*}
\centering
\caption{Observed parameters for the Fimbulthul stream candidate stars as determined by \gaia and the GALAH survey.}
\label{table:star_params}
\begin{tabular}{rcrrrrrrrrrrrrr}
\hline

\texttt{source\_id} & Symbol & $l$ & $b$ & \rv\ & \bprp & G & \teff & \logg & \feh\\
 & & (deg) & (deg) & (\kms) &  &  & (K) & & \\
\hline
6192933650707925376 & $\blacktriangle$ &311.06 & 37.19 & 218.5 & 1.03 & 12.40 & $5154\pm130$ & $2.38\pm0.20$ & $-1.53\pm0.08$\\
6182748015506372480 & $\blacktriangledown$ &308.75 & 32.86 & 214.8 & 1.28 & 12.17 & $4509\pm64$ & $1.24\pm0.17$ & $-1.88\pm0.07$\\
6086940734091928064 & $\blacksquare$ &307.93 & 15.89 & 240.3 & 1.34 & 13.72 & $4774\pm248$ & $1.63\pm0.20$ & $-1.74\pm0.09$\\
6137654604109711360 & \ding{58} &306.18 & 20.92 & 213.5 & 1.31 & 12.83 & $4806\pm73$ & $2.16\pm0.17$ & $-0.46\pm0.07$\\
6140730767063538048 & \ding{54} &305.96 & 21.51 & 261.4 & 1.39 & 13.67 & $4577\pm98$ & $2.13\pm0.20$ & $-0.68\pm0.08$\\
\hline
\end{tabular}
\end{table*}

\begin{figure*}
    \includegraphics[width=\textwidth]{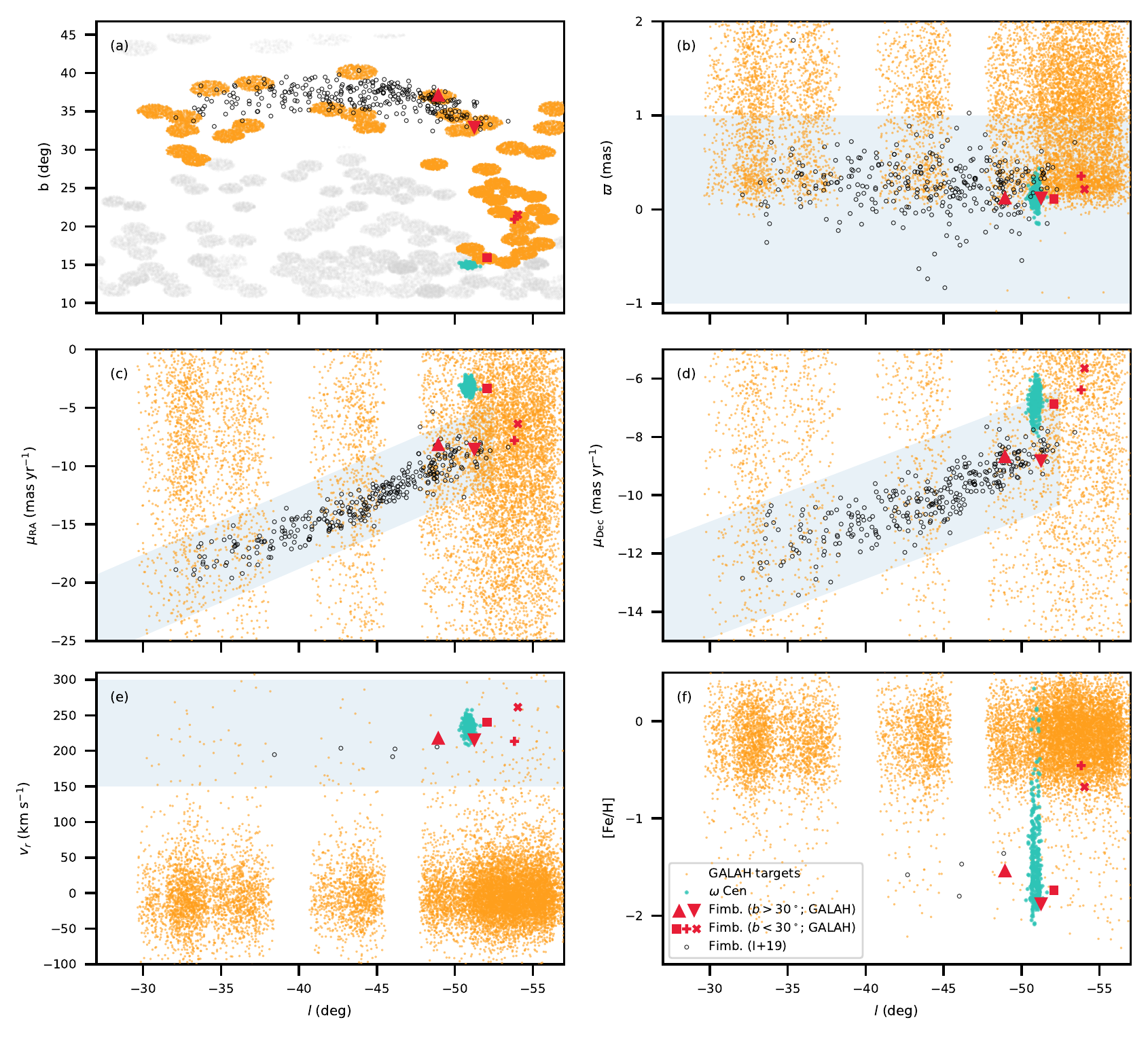}
    \caption{Physical properties of the stars in the region of the sky of \ocen\ and the Fimbulthul stream with respect to the Galactic latitude ($l$). The panels show the: (a) sky distribution of the stars; (b) parallaxes; (c, d) proper motions; (e) radial velocities; and (f) metallicities. In each panel we show the stream members from \citetalias{Ibata2019a} (unfilled black circles), and various groups of stars observed by GALAH: members of \ocen\ (turquoise dots), stars along the stream fiducial on the sky (orange dots), candidate stream members (triangle symbols [$b>30\degr$], and plus, cross \& square symbols [$b<30\degr$]). In the first panel we also show the rest of the stars observed by GALAH in this region of the sky as fainter grey dots. The shaded regions in panels (b, c, d, e) graphically show the selection criteria.}
    \label{fig:member_identification}
\end{figure*}

\begin{figure}
    \includegraphics[width=\columnwidth]{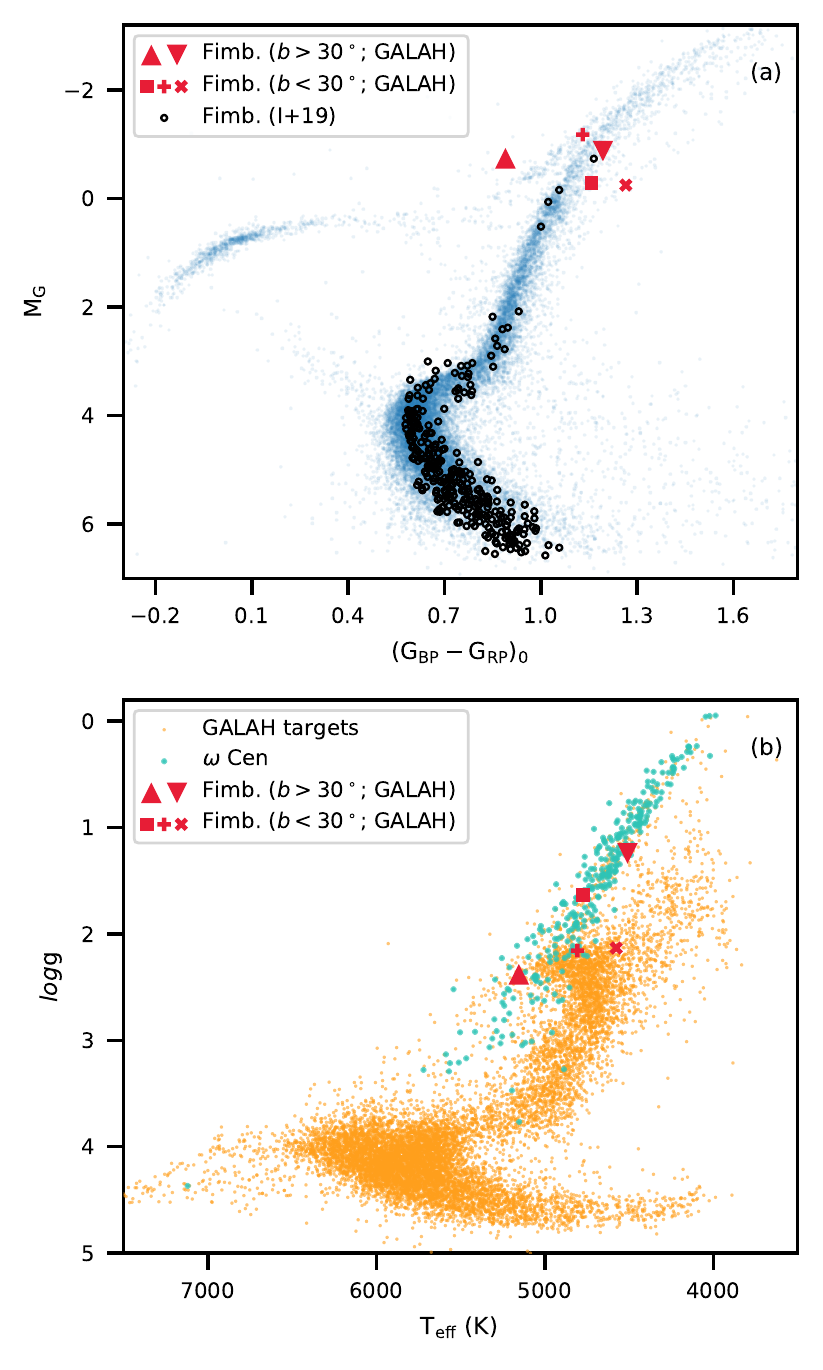}
    \caption{(a) Absolute and dereddened colour-magnitude diagram of \ocen\ in Gaia DR2 photometry, with the stars selected as described in Section~\ref{sec:data}. Over-plotted are the candidate stream members from GALAH (with the same symbols as Figure~\ref{fig:member_identification}; red triangle symbols [$b>30\degr$], and red plus, cross \& square symbols [$b<30\degr$]); and the Fimbulthul stars from \citetalias{Ibata2019a} (open black circles). \response{(b) \teff-\logg diagram of the region of the sky shown in Figure~\ref{fig:member_identification}, highlighting the \ocen and Fimbumthul stars.}
        }
    \label{fig:cmd}
\end{figure}

The proposed Fimbulthul stream members \citepalias[from supplementary table 1 of][]{Ibata2019a} are dominated by dwarf stars with a handful of giant stars. With a distance of $\sim$4~kpc, almost all of the stream stars are much fainter than the typical apparent magnitude range observed by GALAH ($12<V<14$), and none of the proposed stream members were observed by GALAH.

The \textsc{streamfinder} software \citep{Malhan:2018dp} as implemented by \citet{Ibata2019} avoided the region within 30~deg of the Galactic plane due to confusion from the high stellar density and differential reddening. This means that they excluded stars within $\sim15$~deg of \ocen, so we extended our search for possible Fimbulthul stars into this ``zone of avoidance''. We also looked along the known stream region for stars observed by GALAH to identify if there were stars that also shared the kinematics of the known stream stars that may have been missed by \citet{Ibata2019}.

Figure~\ref{fig:member_identification} shows the stars observed by GALAH in the region of the sky around \ocen\ and the Fimbulthul stream, as well as the Fimbulthul members identified by \citetalias{Ibata2019a}. Along the known stream region (i.e., $b>30$~deg), we selected stars observed by GALAH that \response{were from fields with centres} within 4~deg of the fiducial line of the stream and applied the following selection criteria: radial velocity in the range $150<\rv<300$~\kms\ (the radial velocities predicted by \citetalias{Ibata2019a} for the stream) and parallax in the range $|\varpi|<1$~mas. For the proper motion criterion, we noted that the proper motion of the stream is approximately linear with Galactic longitude, so a straight line was fitted to this correlation, and it was required that \pmra\ and \pmdec\ of stars were within 3.4~\masyr\ and 1.9~\masyr\ respectively of these linear fits (these limits were chosen to encompass almost all of the members identified by \citetalias{Ibata2019a}; they are shown as the shaded regions on Figure~\ref{fig:member_identification}b and c). These criteria identified two stars observed by the GALAH survey that are consistent with the sky position and kinematic properties of the Fimbulthul stream, but had not been associated with the stream by \citetalias{Ibata2019a}. Their details (\texttt{source\_id} of \sourceid{6182748015506372480} and \sourceid{6192933650707925376}) are given in Table~\ref{table:star_params} and Table~\ref{table:abund_params} and they are shown as large red triangle symbols on Figures \ref{fig:member_identification}, \ref{fig:cmd}, \ref{fig:abundance_plot}, \ref{fig:abun1_abund2},  and \ref{fig:abundance_plot_1}.

For the zone of avoidance region, we selected the stars observed by GALAH that are along the likely fiducial line of the stream \citepalias[see figure 1 of][]{Ibata2019a}. We used the same radial velocity and parallax criteria as above, but with a proper motion selection of $-10<\pmra<-3$~\masyr, and $-10.0<\pmdec<-5.5$~\masyr, again based upon figure 1 of \citetalias{Ibata2019a}. These criteria identified three stars that could possibly be associated with \ocen\ and the Fimbulthul stream. Their details are given in Table~\ref{table:star_params} and Table~\ref{table:abund_params} and they are shown as the cross, plus, and square symbols on Figures \ref{fig:member_identification}, \ref{fig:cmd}, \ref{fig:abundance_plot}, \ref{fig:abun1_abund2},  and \ref{fig:abundance_plot_1}.

The two stars at the location of the stream (i.e., $b>30$~deg) are metal poor ($\feh<-1.0$), but only one of the zone of avoidance stars (\sourceid{6086940734091928064}) is very metal poor: $\feh=-1.74$. The other two stars are only relatively metal poor: $\feh=-0.46$ and $-0.68$. We explore the abundance properties of the stars further in Section~\ref{sec:chemical_tagging}, but we note here that although their iron abundance is within the observed distribution of \ocen (see Figure~\ref{fig:member_identification}f), these two more metal-rich stars (\sourceid{6137654604109711360}, \sourceid{6140730767063538048}) do not have the expected s-process abundances for their iron abundance to be chemically tagged to \ocen\ \response{\citep[see e.g.,][]{Johnson:2010fs,Marino2011}}. The very metal-poor star is incidentally the star closest to \ocen\ on the sky, less than 1.5~deg from the cluster centre. This places it just outside of the tidal radius of \ocen\ \citep[$\sim1$~deg;][]{DaCosta2008}.

\subsection{Colour-absolute magnitude diagram}
Figure~\ref{fig:cmd} shows the extinction-corrected colour-absolute-magnitude diagram (CMD) of the Fimbulthul stream stars from GALAH and \citetalias{Ibata2019a} stream stars compared to the cluster sequence of \ocen. The photometry has been adjusted to absolute magnitudes and de-reddened colours, assuming $A_V = 3.1\times\ebv$, and the reddening and extinction corrections are from equation 1 and table 1 of \citet{Babusiaux:2018di}. The \ebv\ values for each star were calculated from the \citet{Schlafly:2011iu} reddening maps using the \textsc{dustmaps} interface \citep{Green2018a}, except for the \ocen\ stars, for which we assumed a single reddening value of $\ebv=0.12$ \citep{Harris:1996fr}.

The distance modulus of \ocen\ is well-constrained \citep[e.g.,][]{Braga2018}, and we assumed that \ocen\ and proposed stream members with $b<30$~deg are the same distance, namely $(m-M)_0=13.7\equiv 5.5~\textrm{kpc}$. As found in \citetalias{Ibata2019a}, the stream and \ocen\ are not at the same distance. For the \citetalias{Ibata2019a} sample of Fimbulthul stars, we used a distance modulus of $(m-M)_0=12.9\equiv 3.8~\textrm{kpc}$, i.e., the stream is about 1.7~kpc closer to the Sun than \ocen. We note that \citetalias{Ibata2019a} had a distance modulus difference between the cluster and the stream of 0.7~mag, compared to our 0.8~mag.

The photometry of the bluest of the GALAH stars on the stream (\sourceid{6192933650707925376}) places it on the asymptotic giant branch (AGB) of \ocen. We also inspected CMDs generated from SkyMapper DR2 (Onken et al 2019, in prep) and Pan-STARRS1 DR2 \citep{Chambers:2016vk}, finding the same result, i.e., \sourceid{6192933650707925376} is clearly bluer than the RGB for its luminosity. This is consistent with the GALAH Kiel diagram \response{(Figure~\ref{fig:cmd}b)}, which identifies it as an AGB star from its \teff\ and \logg.

\section{Chemical tagging}\label{sec:chemical_tagging}

\begin{table*}
\centering
\caption{Fimbulthul stream candidate stars observed by the GALAH survey. Elemental abundances with blank entries were those that did not meet the \texttt{flag\_x\_fe} criteria defined in Section \ref{sec:data}. The flag values are given in Table \ref{table:flag_params}. The online version contains all of the \ocen\ stars observed by GALAH, and columns for all the elemental abundances.}
\label{table:abund_params}
\begin{tabular}{rcrrrrrrrrrrrrr}
\hline

\texttt{source\_id} & Symbol & \feh & \alphafe & \yfe & \bafe & \nafe & \alfe \\
 &   &  &  &  &  &  &  \\
\hline
6192933650707925376 & $\blacktriangle$ &$-1.53\pm0.08$ &  & $0.91\pm0.09$ & $1.18\pm0.11$ & $0.29\pm0.05$ & $0.58\pm0.05$\\
6182748015506372480 & $\blacktriangledown$ &$-1.88\pm0.07$ & $0.18\pm0.03$ & $-0.03\pm0.08$ & $-0.09\pm0.09$ & $0.23\pm0.05$ & $0.38\pm0.04$\\
6086940734091928064 & $\blacksquare$ &$-1.74\pm0.09$ & $0.28\pm0.04$ & $0.26\pm0.10$ & $0.77\pm0.11$ &  & \\
6137654604109711360 & \ding{58} &$-0.46\pm0.07$ & $0.22\pm0.02$ & $0.19\pm0.08$ & $0.28\pm0.09$ & $0.15\pm0.05$ & $0.33\pm0.04$\\
6140730767063538048 & \ding{54} &$-0.68\pm0.08$ & $0.28\pm0.03$ & $-0.06\pm0.09$ & $-0.20\pm0.11$ & $0.20\pm0.06$ & $0.43\pm0.05$\\
\hline
\end{tabular}
\end{table*}

\begin{figure}
    \includegraphics[width=\columnwidth]{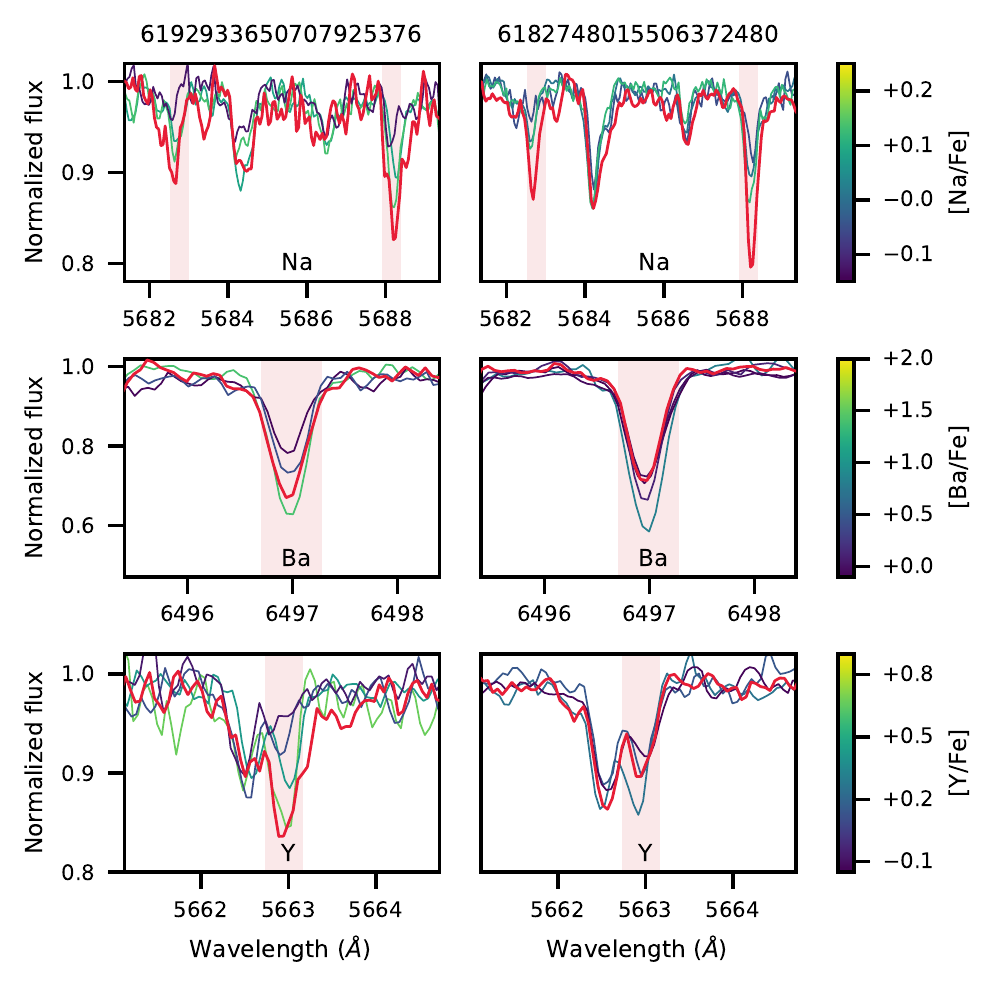}
    \caption{\response{Continuum normalized spectra for \protect\sourceid{6192933650707925376} (left column) and \protect\sourceid{6182748015506372480} (right column) highlighting the bandpasses used for determining the elemental abundances for Na (top row), Ba (middle row), and Y (bottom row) in GALAH DR2. In each panel are the spectra of stars from the GALAH DR2 dataset with \teff, \logg, and \feh\ within $\pm100$~K, $\pm0.5$~dex, $\pm0.03$~dex respectively of our Fimbulthul stream stars, but with a range of abundances in the given element. These spectra are colour coded by their [X/Fe]. These plots show that the two stream stars have stronger Na lines than stars of the same of stellar parameters, while \protect\sourceid{6192933650707925376} has strong Ba and Y lines and \protect\sourceid{6182748015506372480} has weaker lines for these elements. }}
    \label{fig:spec_regions}
\end{figure}

\begin{figure*}
    \includegraphics[width=\textwidth]{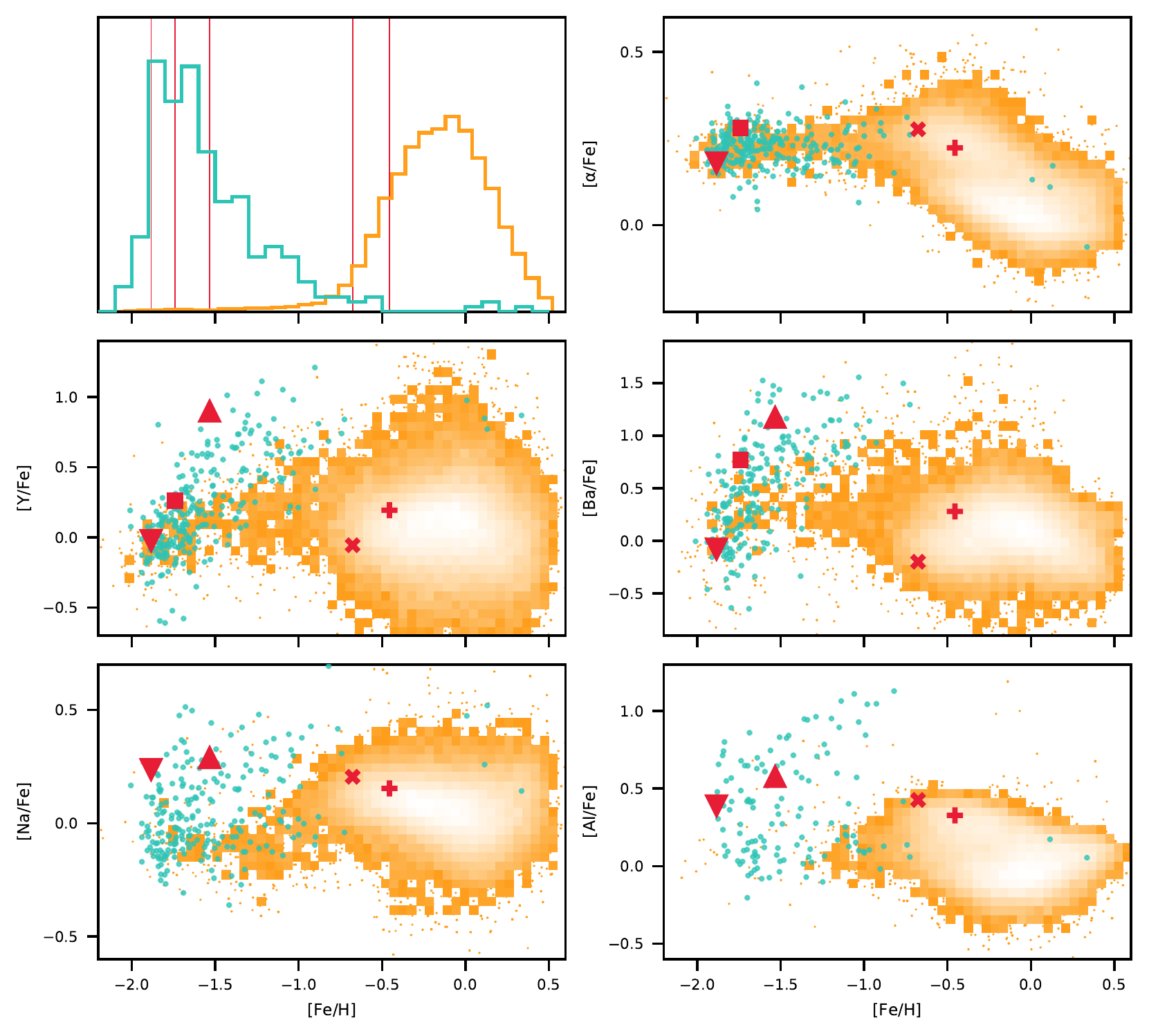}
    \caption{Distributions of (top-left panel) iron abundance, and (remaining panels) \alphafe, \yfe, \bafe, \nafe, and \alfe\ against \feh\ as determined by the GALAH survey for stars in the region of the sky around \ocen\ and the candidate Fimbulthul stream members identified in this work. The top-left panel shows the normalized iron abundance distributions of the cluster (turquoise distribution) and field (orange distribution). The vertical red lines mark the iron abundances of our candidate stream members. In the remaining panels is a log-scaled colour density map, with the lightest squares containing the most stars (truncated at 3 stars per bin; below that threshold, individual stars are plotted as orange dots). The distributions for \ocen\ (turquoise dots) and the candidate Fimbulthul stream stars observed by GALAH (same symbols as in Figure~\ref{fig:member_identification}; red triangle symbols [$b>30\degr$], and red plus, cross \& square symbols [$b<30\degr$]) are overlaid. For \alphafe, we find a distribution of \ocen like that of other metal-poor stars. For \nafe\ and \alfe\ we find the characteristic star-to-star abundance dispersions previously observed in \ocen, and that the two GALAH-Fimbulthul stars (red triangle symbols) are enhanced in these elements. In the s-process elements we find the well-known correlation between iron abundance and s-process elemental abundance, with one of the GALAH-Fimbulthul stars very enhanced in yttrium and barium. For the stars from the zone of avoidance region, it is likely that the two metal-rich stars (red plus and cross) are not related to \ocen due in particular to their s-process abundances.}
    \label{fig:abundance_plot}
\end{figure*}

\begin{figure}
    \includegraphics[width=\columnwidth]{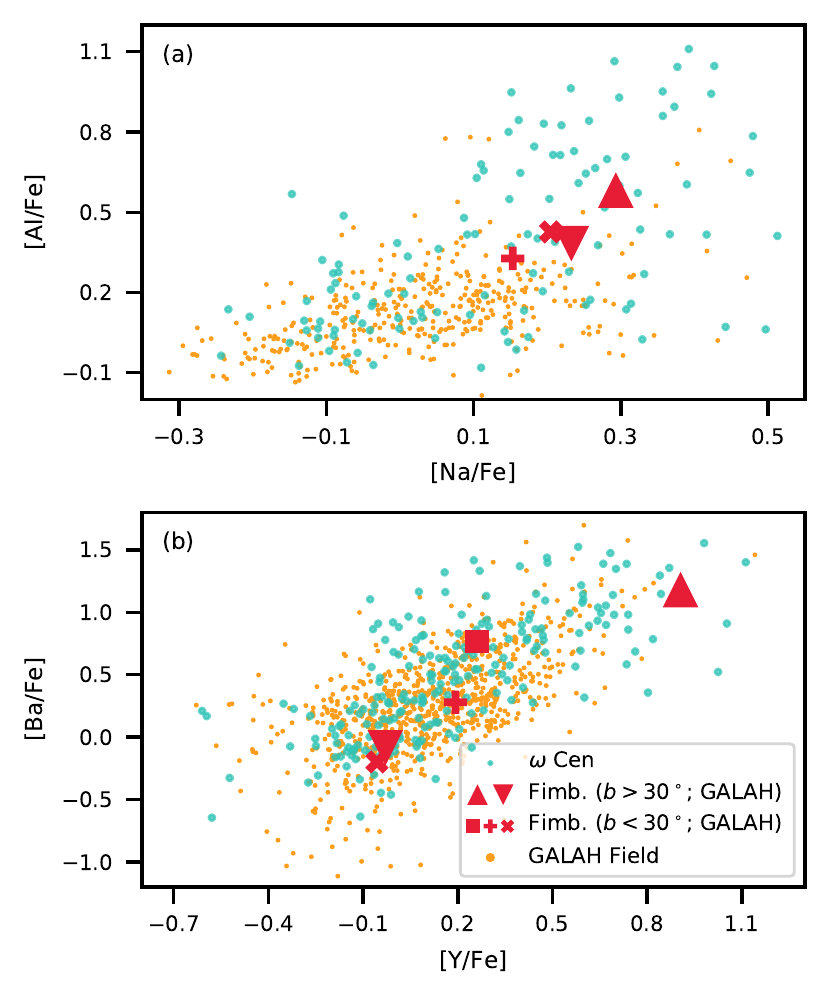}
    \caption{Abundance distributions of \ocen\ compared to all other metal-poor stars ($\feh<-0.8$) observed by GALAH in the region of the sky shown in Figure~\ref{fig:member_identification}. (a) \nafe\ against \alfe\ (i.e., light elements); and (b) \yfe\ against \bafe\ (i.e., s-process elements). About half of the \ocen\ stars (turquoise dots) and the two candidate Fimbulthul stars with $b>30$~deg (red triangles) have enhanced light elements compared to the bulk of the other metal-poor stars. Most of the field stars have $\alfe\sim0$, while the \ocen\ stars extend up to $\alfe\sim1$. For the s-process elements, there is a large range in the field, but the one candidate Fimbulthul star is even more enhanced in \yfe\ than the majority of field stars.}
    \label{fig:abun1_abund2}
\end{figure}

As discussed in Section~\ref{sec:intro}, globular clusters, and in particular \ocen, display star-to-star abundance patterns that are not observed in the vast majority of field stars. Relevant to this work, the elemental abundances of stars in \ocen have been found to cover large ranges in light elements (e.g. sodium\footnote{Although oxygen is typically part of the analysis of second population stars in globular clusters, the wavelength coverage of HERMES does not include useful oxygen lines in cool, metal-poor giants. It would therefore be useful to follow up these stars (e.g., in the infrared) to obtain oxygen abundances.} and aluminium). In addition, unlike almost every other globular cluster, the stars also exhibit a large range of iron abundances ($-2.0<\feh<-0.5$) and the abundances of s-process elements \citep[e.g., lanthanum, yttrium, and barium; see e.g.,][]{Norris1995a,Norris1995,Stanford:2010fo,Johnson:2010fs,Marino:2012eg,Simpson:2013ee}. These provide useful markers for chemically tagging stars to \ocen.

In Figure~\ref{fig:abundance_plot} we show the iron abundance, \alphafe, and abundance distributions for \yfe, \bafe, \nafe, and \alfe\ for the 40,040 stars in the region of the sky within the box $10^\circ<b<45^\circ$ and $-27^\circ<l<-57^\circ$ (i.e., the region of the sky shown in Figure~\ref{fig:member_identification}a). The abundances of the \ocen stars and our proposed Fimbulthul stream members observed by GALAH are highlighted. The full abundance set is shown in Figure~\ref{fig:abundance_plot_1}. \response{In Figure~\ref{fig:spec_regions} we show small regions of the HERMES spectrum surrounding the regions used by \textsc{the cannon} for Na, Ba and Y abundance determination. Spectra of the two new Fimbulthul stream stars are shown with heavier red lines, and spectra of other GALAH stars with similar stellar parameters but different elemental abundances are shown with thin lines colour-coded by their abundance. The higher Ba and Y abundances in \sourceid{6192933650707925376} and the enhanced Na abundance in both stars can be seen directly in the spectra.}

Overall, our results show that we recover the known iron abundance distribution of \ocen\footnote{Four of the \ocen\ stars have implausible, super-solar iron abundances. In the provisional internal third data release of GALAH, these stars have much more reasonable metallicities of $\feh\sim-0.7$, while the other \ocen\ stars are relatively unchanged.}, with most of the stars much more metal poor than the bulk of the stars in this region of the sky. For the light elements \nafe\ and \alfe, we find the large star-to-star abundance ranges known to exist in \ocen \citep[e.g.,][]{Norris1995a}. The two stars identified in the Fimbulthul region (i.e., $b>30$~deg) also are enhanced in these elements, strongly indicative of them being second population globular cluster stars. In Figure~\ref{fig:abun1_abund2}a, we highlight the correlation between \nafe\ and \alfe\ for all stars with $\feh<-0.8$ observed by GALAH in this region of the sky. The bulk of the field stars have \alfe\ and \nafe\ less than 0.4~dex, but many of the \ocen\ stars are enhanced in these elements like the two GALAH-Fimbulthul stars.

Further evidence for chemically tagging these stars to \ocen\ can be found in their s-process abundances. The \ocen\ stars show the characteristic rapid increase in s-process abundance with iron found in other work.  The star that is more metal rich (\sourceid{6192933650707925376}) has $\yfe=0.9$ and $\bafe=1.2$. Figure~\ref{fig:abun1_abund2}b shows that such abundances are rarely seen in the field. Finding a metal-poor star that is both enhanced in s-process elements and light elements like aluminium is rare. There are $\sim20$ stars with $\feh<-1.0$ in the entire GALAH dataset with $\bafe>1.0$ and $\alfe>0.4$ and most are found in \ocen. The other stars with enhanced light element abundances warrant follow up to understand their origins. The more metal-poor star (\sourceid{6182748015506372480}; the reddest GALAH-Fimbulthul star on Figure~\ref{fig:cmd}) shows no enhancement in \yfe\ nor \bafe\ relative to the field, but this would be expected at its iron abundance if it had been lost from \ocen.

Both \sourceid{6182748015506372480} and \sourceid{6192933650707925376} have second population abundances, they follow the \feh-[s/Fe] trend of \ocen, and they are spatially and kinematically compatible with being members of the Fimbulthul stream. As a result, we are confident about chemically tagging these stars to \ocen, and therefore the rest of the Fimbulthul stream to \ocen.

We now briefly consider the three stars from the zone of avoidance \citep[i.e., $b<30$, outside of the region considered by][]{Ibata2019}. \response{The abundances of two metal-rich stars are not consistent with their being former members of} \ocen. Although somewhat enhanced in light elements, they do not show the level of enhancement in s-process elements expected for an \ocen\ star of their iron abundance \response{\citep[see e.g.,][]{Johnson:2010fs,Marino2011}}. Unfortunately, the extra-tidal star closest to \ocen\ did not have reliable light element abundances, but it does have a moderate enhancement in its s-process abundances. Its closeness to the cluster (only 1.5~deg away), and kinematics basically identical to rest of the cluster point to it being a relatively recently escaped extra-tidal star. Its position on the sky aligns with the direction of the extra-tidal features found by \citetalias{Ibata2019a}.

\subsection{Abundance flagging}

\begin{table*}
\centering
\caption{Stellar parameter and elemental abundance flagging for the five stars of interest from the GALAH survey. Values of zero indicate there are known reliability problems with the value. $+1$ flags that $\textsc{the cannon}$ had started to extrapolote; $+2$ flags that the line strength was below the 2-$\sigma$ upper limit. For 6086940734091928064, the values for \nafe\ and \alfe\ (e.g., $\geq4$) show that this elemental abundance had multiple reliability problems.}
\label{table:flag_params}
\begin{tabular}{rcrrrrrrrrrrrrr}
\hline

\texttt{source\_id} & \texttt{sobject\_id} & Symbol & \texttt{flag\_cannon} & \texttt{flag\_y\_fe} & \texttt{flag\_ba\_fe} & \texttt{flag\_na\_fe} & \texttt{flag\_al\_fe} \\
\hline
6192933650707925376 & 170512001301266 & $\blacktriangle$ &0 & 1 & 1 & 1 & 2\\
6182748015506372480 & 170513003501228 & $\blacktriangledown$ &0 & 0 & 2 & 1 & 1\\
6086940734091928064 & 170404002101197 & $\blacksquare$ &0 & 0 & 0 & 4 & 5\\
6137654604109711360 & 170408004001066 & \ding{58} &0 & 0 & 0 & 0 & 0\\
6140730767063538048 & 170408004001278 & \ding{54} &0 & 0 & 0 & 0 & 0\\
\hline
\end{tabular}
\end{table*}

As discussed in Section~\ref{sec:data}, we have included stars in our sample that were flagged as having potentially unreliable abundance values. In Table~\ref{table:flag_params} we give the \texttt{flag\_cannon} and \texttt{flag\_x\_fe} values for the five stars highlighted throughout this work. Of note is that the two likely Fimbulthul stream stars (\sourceid{6182748015506372480} and \sourceid{6192933650707925376}) both have their \bafe, \nafe, and \alfe\ abundances flagged as either 1 (\textsc{the cannon} started to extrapolate) or 2 (Line strength below 2-$\sigma$ upper limit). In the case of the $\texttt{flag\_x\_fe}=1$, this occurs because the training set does not include metal-poor stars with abundances like those seen in second population globular cluster stars. As such, \textsc{the cannon} needs to extrapolate in order to estimate abundances. This does not necessarily mean that the values are inaccurate; moreover abundance accuracy is not the key aim of this work. Instead, our goal was to identify stars with enhanced abundances of light elements relative to the bulk of the metal-poor halo population, in order to chemically tag them as second population \ocen stars. We can be confident that \textsc{the cannon} will have estimated abundances that reflect these stars are enhanced in the particular elements of interest.

\response{\subsection{Abundances on the asymptotic giant branch}
As noted in Section \ref{sec:stream_ident}, \sourceid{6192933650707925376} is an asymptotic giant branch (AGB) star. There has been much discussion in the literature about differences in the abundances of stars on the AGB versus those on the RGB in globular clusters \citep{Norris:1981ij,Campbell:2013bd,Campbell2017,MacLean:2016gv,MacLean2018}, in particular with respect to apparent differences in their derived metallicities and light element abundances \citep{Lapenna:2016ha,Mucciarelli2019}. There are two related effects that are causing these discrepancies. First, the iron abundance derived from neutral and ionized species differ in AGB stars. Second, on the AGB, the \teff derived from the infrared flux method gives significantly different values to those derived from spectroscopic analysis --- while in RGB stars, the two methods give the same values. These discrepancies are likely the result of the current treatment of stellar atmospheres using 1D-LTE, and the potential for  3D and/or NLTE effects to be larger in AGB stars than in RGB stars. For GALAH DR2 abundances there was an effort to include non-LTE calculations for key elements, including Na and Al \citep{Buder2018}.}

\section{Concluding remarks}\label{sec:discussion}
In this work, we have shown that two stars observed by the GALAH survey are associated with the Fimbulthul stellar stream, and can be chemically tagged to the massive globular cluster \ocen. We find another star that has likely just escaped from the cluster. \ocen\ is the most massive of the Milky Way's $\sim160$ known clusters and has a broad iron distribution, abundance patterns not otherwise seen in globular clusters, and a retrograde orbit about the Galaxy. This has led to the hypothesis that \ocen\ is the core of an accreted dwarf galaxy. The progenitor galaxy of \ocen\ would have been $10^3$ to $10^4$ times larger than the present day mass of \ocen\ \citep{Tsuchiya2003,Tsuchiya2004}, which means that there should be tidal tails and debris lost from \ocen\ and its host galaxy throughout the Milky Way. Confirming the association of Fimbulthul to \ocen\ by \citetalias{Ibata2019a} marks the first time that tidal arms have been found well away from the cluster.

Overall, we have only identified three stars observed by GALAH, outside of the tidal radius that we are confident to associate with \ocen.  As discussed in \citet{DaCosta2008}, the current orbit of \ocen is such that it is strongly influenced by tidal shocks. Stars leaving the cluster with a relative velocity of $\sim1$~\kms\ \citep[the velocity dispersion in the outer parts is a few \kms; see, e.g., ][]{DaCosta:2012hd} can transverse the region between 1 and 2 tidal radii in less than the orbital period around the Galaxy. At this point the next disk crossing likely gives them sufficient kick to move further away. As such, we would expect there to be few stars remaining close to cluster --- there is effectively just too much gravitational energy being pumped in. Note that this is quite different to the situation for an orbit with a larger ratio of Galactic apocentre to pericentre, in which the unbound stars do not dissipate as rapidly and we would expect tidal debris near the cluster.

In the next few years, both the \gaia and GALAH surveys should produce their third data releases, which will add a vast amount of new and improved data by which our study could be expanded. Having demonstrated here the power of chemical tagging in aiding the identification of stellar streams from \ocen, it will be interesting to see whether those new and expanded data sets allow us to further strengthen the connection between the Fimbulthul stream and \ocen, and to potentially draw conclusions on the timescales on which the disruption of the postulated progenitor galaxy by the Milky Way has occurred.

\section*{Acknowledgements}

The GALAH survey is based on observations made at the Anglo-Australian  Telescope, under programmes A/2013B/13, A/2014A/25, A/2015A/19, A/2017A/18. We acknowledge the traditional owners of the land on which the AAT stands, the Gamilaraay people, and pay our respects to elders past and present. \response{This paper includes data that have been provided by AAO Data Central (\url{datacentral.org.au}).}

\response{We thank the referee for their helpful comments that improved the manuscript, in particular regarding AGB stars.}

The following software made this research possible: \textsc{configure} \citep{Miszalski:2006ef}; \textsc{iraf} \citep{Tody:1986df}; \textsc{python} (version 3.7.4); \textsc{matplotlib} \citep[v3.1.1][]{Hunter:ih,Caswell2019} \textsc{astropy} \citep[version 3.2.2;][]{TheAstropyCollaboration:2013cd,TheAstropyCollaboration:2018ti}, a community-developed core \textsc{python} package for Astronomy; \textsc{pandas} \citep[version 0.25.1;][]{McKinney:2010un}; \textsc{topcat} \citep[version 4.6-3;][]{Taylor:2005wx}.

This work has made use of data from the European Space Agency (ESA) mission {\it Gaia} (\url{https://www.cosmos.esa.int/gaia}), processed by the {\it Gaia} Data Processing and Analysis Consortium (DPAC, \url{https://www.cosmos.esa.int/web/gaia/dpac/consortium}). Funding for the DPAC has been provided by national institutions, in particular the institutions participating in the {\it Gaia} Multilateral Agreement.

Parts of this research were conducted by the Australian Research Council Centre of Excellence for All Sky Astrophysics in 3 Dimensions (ASTRO 3D), through project number CE170100013. JDS, SLM and DZ acknowledge the support of the Australian Research Council through Discovery Project grant DP180101791. SB acknowledges funds from the Alexander von Humboldt Foundation in the framework of the Sofja Kovalevskaja Award endowed by the Federal Ministry of Education and Research. TZ acknowledge the financial  support  from  the  Slovenian  Research  Agency  (research core funding No. P1-0188). KF and Y-ST are grateful for support from Australian Research Council grant DP160103747. Y-ST is grateful to be supported by the NASA Hubble Fellowship grant HST-HF2-51425.001 awarded by the Space Telescope Science Institute.





\section*{Affiliations}
{\small\it
\noindent
$^{1}$School of Physics, UNSW, Sydney, NSW 2052, Australia\\
$^{2}$Centre of Excellence for Astrophysics in Three Dimensions (ASTRO-3D), Australia\\
$^{3}$Research School of Astronomy \& Astrophysics, Australian National University, ACT 2611, Australia\\
$^{4}$Centre for Astrophysics, University of Southern Queensland, Toowoomba, QLD 4350, Australia\\
$^{5}$Department of Physics and Astronomy, Johns Hopkins University, Baltimore, MD 21218, USA\\
$^{6}$Institute for Advanced Study, Princeton, NJ 08540, USA\\
$^{7}$Department of Astrophysical Sciences, Princeton University, Princeton, NJ 08544, USA\\
$^{8}$Observatories of the Carnegie Institution of Washington, 813 Santa Barbara Street, Pasadena, CA 91101, USA\\
$^{9}$Sydney Institute for Astronomy, School of Physics, A28, The University of Sydney, NSW, 2006, Australia\\
$^{10}$Max Planck Institute for Astronomy (MPIA), Koenigstuhl 17, 69117 Heidelberg, Germany\\
$^{11}$Australian Astronomical Optics, Faculty of Science and Engineering, Macquarie University, Macquarie Park, NSW 2113, Australia\\
$^{12}$Faculty of Mathematics and Physics, University of Ljubljana, Jadranska 19, 1000 Ljubljana, Slovenia\\
$^{13}$Department of Physics and Astronomy, Uppsala University, Box 517, SE-751 20 Uppsala, Sweden\\
$^{14}$Department of Physics and Astronomy, Macquarie University, Sydney, NSW 2109, Australia\\
$^{15}$School of Physical and Chemical Sciences, University of Canterbury, New Zealand\\
$^{16}$Monash Centre for Astrophysics, School of Physics and Astronomy, Monash University, Australia\\
}



\appendix

\section{Full abundance set}

\begin{figure*}
    \includegraphics[width=\textwidth]{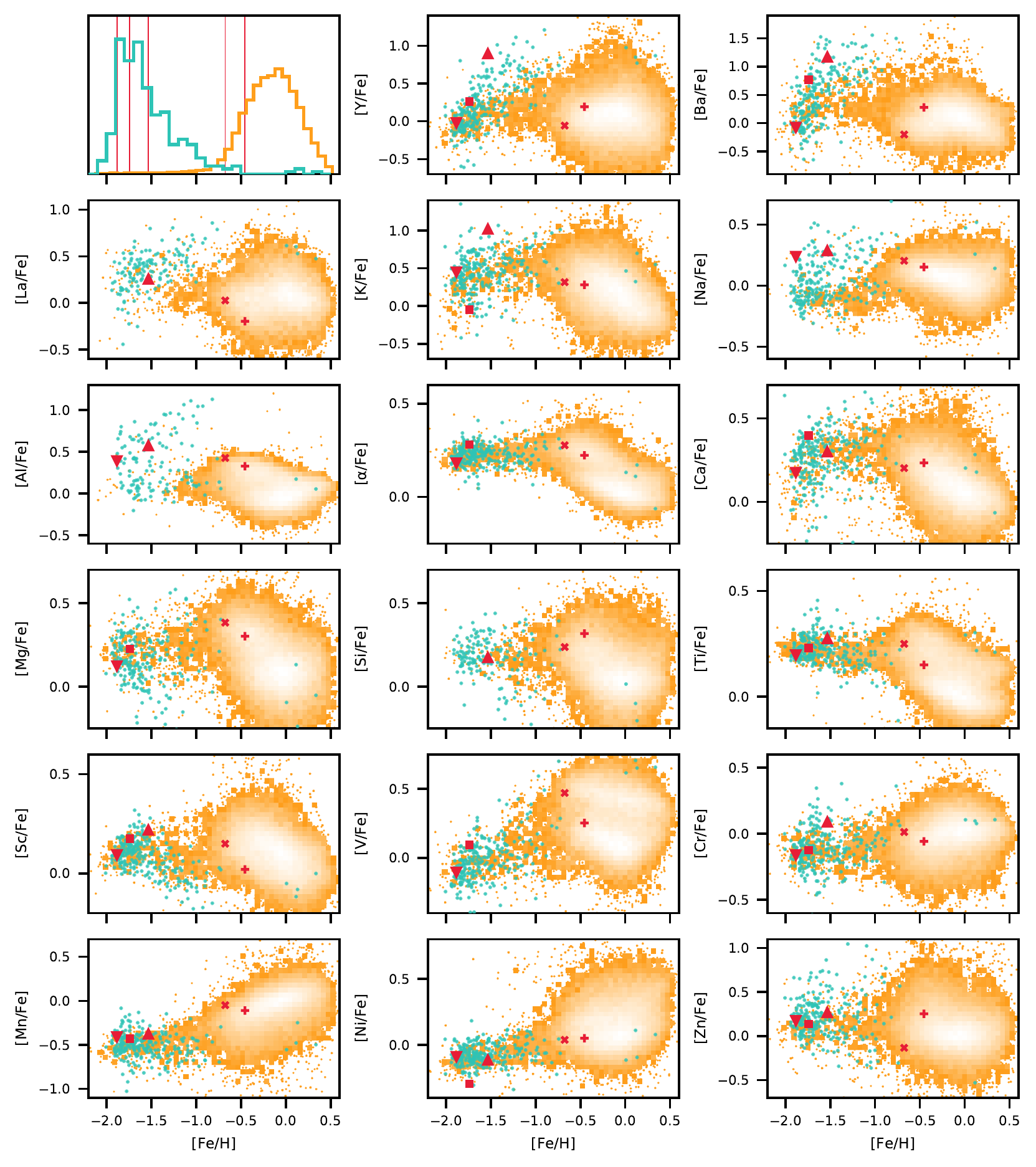}
    \caption{The same as Figure~\ref{fig:abundance_plot} but for the entire set of elemental abundances available within the GALAH catalogue.}
    \label{fig:abundance_plot_1}
\end{figure*}


\bsp	
\label{lastpage}
\end{document}